\documentclass[twocolumn,pre,showpacs,superscriptaddress]{revtex4}

\usepackage{epsfig}

\newcommand{\ra}{\langle r\rangle}

\begin{document}

\title{Self-organization with equilibration:\\
 a model for the intermediate phase in rigidity percolation}

\author{M.V.~Chubynsky}
\email{mykyta.chubynsky@umontreal.ca}
\affiliation{
        D\'epartement de Physique,
	Universit\'e de Montr\'eal,
       C.P. 6128, Succursale Centre-ville, Montr\'eal,
        Qu\'ebec, Canada H3C 3J7}
\author{M.-A.~Bri\`ere}
\email{marc-andre.briere@umontreal.ca}
\affiliation{
        D\'epartement de Physique,
	Universit\'e de Montr\'eal,
       C.P. 6128, Succursale Centre-ville, Montr\'eal,
        Qu\'ebec, Canada H3C 3J7}
\author{N. Mousseau}
\email{normand.mousseau@umontreal.ca}
\affiliation{
        D\'epartement de Physique,
	Universit\'e de Montr\'eal,
       C.P. 6128, Succursale Centre-ville, Montr\'eal,
        Qu\'ebec, Canada H3C 3J7}
\affiliation{Institute for Theoretical Physics,
  Universiteit Utrecht, Leuvenlaan 4, 3584 CE
  Utrecht, the Netherlands}

\date{\today}

\begin{abstract}
  Recent experimental results for covalent glasses suggest the
  existence of an intermediate phase attributed to the
  self-organization of the glass network resulting from the tendency
  to minimize its internal stress.  However, the exact nature of this
  experimentally measured phase remains unclear. We modify a
  previously proposed model of self-organization by generating a
  uniform sampling of stress-free networks.  In our model, studied on a diluted
  triangular lattice, an unusual intermediate phase appears, in which
  both rigid and floppy networks have a chance to occur, a result also
  observed in a related model on a Bethe lattice by Barr\'e \textit{et
  al.} [Phys. Rev. Lett. \textbf{94}, 208701 (2005)]. Our results for the
  bond-configurational entropy of self-organized networks, which turns
  out to be only about 2\% lower than that of random networks, suggest that
  a self-organized intermediate phase could be common in systems near
  the rigidity percolation threshold.
\end{abstract}

\pacs{05.65.+b, 65.40.Gr, 61.43.Bn, 64.70.Pf}

\maketitle

\section{Introduction}

Rigidity theory \cite{thorpe83,fengsen,jacobs95,duxbury95,travthorpe}
has improved considerably our understanding of the structural, elastic
and dynamical properties of systems such as chalcogenide glasses
\cite{thorpe83,travthorpe,travbool}, interfaces \cite{lucvac99} and
proteins \cite{pnas02}, as a function of their connectivity. In its
classical version, it introduces the concept of a rigidity transition,
separating a soft (or \textit{floppy}) and a \textit{rigid} phases characterized by a
mean coordination number. In many systems, optima or thresholds of
various physical quantities are often observed at the rigidity
transition.  Some time ago, for example, Phillips \cite{phillips79}
noted that among chalcogenides, the best glassformers have
a mean coordination such that the number of degrees of freedom is equal to
the number of covalent (bond-stretching and bond-bending)
constraints. At this point, corresponding to the rigidity transition,
networks are largely rigid but stress-free. This prevents crystallization
for both kinetic and thermodynamic reasons: being stress-free, the glassy
state is not too energetically unfavorable compared to the crystal; being
rigid, the networks lack the flexibility to efficiently explore the phase
space and reach the crystalline state fast.
Similarly, in the last 5 years, it has become clear that most proteins
in their native state sit almost exactly at the rigidity transition,
which could be necessary to have enough flexibility to fullfill their
function while retaining their overall structure \cite{pnas02}.

Recently, a series of experiments on chalcogenide and oxide glasses
\cite{bool99sise,
bool00sise,bool00asse,bool00geasse,bool01gese,
bool02cr,bool03pse,christova,bool04geps,kostyaPRL,bool05pgese,
bool05geass,bool05geass2,bool05press,bool05sil,sharma} have demonstrated the existence
of a number of interesting and surprising behaviors. For instance,
glasses with nearly optimal properties, such as the absence of aging
\cite{bool04geps,bool05pgese,bool05geass2}  and vanishing
non-reversing enthalpy of the glass transition \cite{bool99sise,
bool00sise,bool00asse,bool00geasse,bool02cr,bool03pse,bool04geps, bool05pgese,
bool05geass, bool05geass2} are observed not just at a particular mean
coordination, but in some range of coordinations, suggesting the
presence of an \textit{intermediate phase} between the floppy and the
rigid phases. While the details and  exact origin of this intermediate phase are
still a matter of debate (see, for example, the recent experimental
paper by Golovchak \textit{et al.} \cite{golovchak06}), it appears that it is due to the
self-organization of the network minimizing the internal stress. If,
according to Phillips' argument, we expect ``optimal'' glasses to be
rigid but stress-free, we should now expect to find this property
everywhere in the intermediate phase rather than only at a single
critical point, as in the standard phase diagram of rigidity percolation;
there is now some direct evidence for this \cite{christova,bool05press}.

A few models were proposed to explain this self-organization. Thorpe
\textit{et al.} \cite{thorpe00, czech01} have shown in an
out-of-equilibrium model that it is possible to generate a stress-free
intermediate phase.  Barr\'e \textit{et al.}  \cite{barre} have 
shown in addition that such a phase was thermodynamically stable on a ring-free
Bethe lattice, where each site has 3 degrees of freedom, but the
whole network is embedded in an infinite-dimensional space. Taking a
different approach, Micoulaut \cite{micoulaut02} demonstrated that one
could recover an intermediate phase by concentrating all the strain in
local structures.

The goal of this paper is to assess whether or not a self-organized
network with a finite-dimensional topology is also thermodynamically
possible. This verification is important on two counts: (1) the Bethe
lattice is a loopless structure producing a first-order rigidity
transition \cite{dux97,dux99,travthorpe} while 2- and 3-dimensional
regular networks undergo a second-order phase transition
\cite{jacobs95,travthorpe}; (2) the original model of Thorpe \textit{et
al.} is an out-of-equilibrium model which could lead to highly
atypical networks.

For simplicity, we study two-dimensional central-force (2D CF)
networks. This allows us to consider networks of much bigger linear
dimensions than is possible in 3D, limiting the finite-size effects.
Moreover, in all cases studied until now, rigidity results for 2D CF
networks have been qualitatively very similar to those obtained for 3D
glass networks. Our results should therefore also apply to glass networks.

In the next section, we review the basic facts about rigidity and
the intermediate phase. We
then explain the model used here and present our results for a bond-diluted
two-dimensional triangular network. We first discuss an unusual property of
our model: a network in the intermediate phase can be either rigid or floppy
with a finite probability. We then focus on the calculation of the entropy of
self-organized networks.

\section{The intermediate phase in the
rigidity phase diagram}

In rigidity theory, an elastic network is characterized by the number
of motions, called \textit{floppy modes}, that do not distort any
constraints.  In a system with no constraints, all degrees of freedom
are floppy modes and thus their number is $dN$, where $N$ is the
number of atoms and $d$ is the dimensionality of space. In an
approximation due to Maxwell and known as Maxwell counting
\cite{maxwell}, it is assumed that each additional constraint removes
a floppy mode, so that the number of floppy modes for a given number
of constraints $N_c$ can be written as
\begin{equation}
F=dN-N_c.\label{Max}
\end{equation}
When the number of constraints becomes equal to the number of degrees
of freedom, $F=0$ and the network undergoes a rigidity percolation
transition, going from floppy to rigid. In a 2D CF network, the number
of constraints per atom is $\ra/2$, where $\ra$ is the mean
coordination (the average number of connections of a site); the
critical coordination is therefore $\ra=4$. In chalcogenide glasses, characterized by the
chemical formula A$_x$B$_y$C$_{1-x-y}$, where A is an atom of valence
4 (usually Ge or Si), B is an atom of valence 3 (As or P) and C is a
chalcogen (Se, S or Te), counting both covalent bonds and their
related angular constraints, the critical coordination is $\ra=2.4$
\cite{thorpe83}.

Since Maxwell counting is a mean-field theory, it ignores fluctuations
and correlations that can be built in the network. An overall rigid
network can have some internal localized floppy modes. Also, a
constraint inserted into a piece of the network that is already rigid
does not remove floppy modes. Such constraints, known as \textit{
redundant}, are obviously present in the rigid phase, where
Eq.~(\ref{Max}) gives a negative number of floppy modes, since this number
cannot be lower than $d(d+1)/2$ (6 in 3D,
3 in 2D), to account for rigid body translations and rotations; but
redundant constraints can also be present, of course, in an overall floppy network. In
generic networks, such as glasses, where the bond lengths vary, these
constraints create an internal stress and increase the elastic energy, since
at least some part of the network has to be deformed to accommodate
them.

If the number of redundant constraints, $N_R$, is known, then the Maxwell
counting formula can be corrected:
\begin{equation}
F=dN-N_c+N_R.\label{Maxexact}
\end{equation}
This result is exact --- the problem lies in calculating $N_R$.  A theorem by
Laman \cite{laman} makes this possible.  Consider all possible subnetworks of a system. If
the number of degrees of freedom minus the number of constraints is less than
$d(d+1)/2$ for at least one subnetwork, there must be redundant
constraints. Laman showed that this is the only way redundant constraints can
occur: for them to be present, the above \textit{must} be satisfied for at least
one subnetwork. This is strictly true in 2D; although there are known
counterexamples for general networks in 3D, it is \textit{assumed} true
(there is no rigorous mathematical proof, but no known counterexamples
either) for glassy networks with covalent bonding including angular
constraints \cite{whiteley84,travwhite}.

Laman's theorem is used in a computer algorithm for rigidity analysis,
known as the \textit{pebble game} \cite{jacobs95,jacobs97,jacobs98,travbio,
proteins01}. The pebble game starts with an empty
network and then one constraint is added at a time and each added
constraint is checked for redundancy. Thus at every stage in the
network-building process the number of redundant constraints, $N_R$,
and, according to Eq.~\ref{Maxexact}, $F$, are known
exactly. Independent constraints are matched to \textit{pebbles} that are
assigned to sites and whose total number is equal to the number of
degrees of freedom; thus the number of free pebbles (not matched to
any constraint) gives the number of floppy modes.  The pebble game, in
addition, identifies stressed regions in the network and also offers
\textit{rigid cluster decomposition} that identifies all rigid clusters
in the network. The analysis provided by the pebble game is purely
topological, the details of the geometry are not taken into account,
nor is the exact expression for the forces. The downside is that the
pebble game (as well as Laman's theorem itself) is only applicable to
\textit{generic} networks: networks special in some way (having parallel
bonds, for instance) may have rigidity properties that are different
from those of the vast majority of networks of a given topology. The
fraction of such special (or \textit{non-generic}) networks among all
networks of a given topology is, however, zero, and so a covalent
glass network, being disordered, can be safely assumed to be
generic. 

Recently, a series of experiments \cite{bool99sise,
bool00sise,bool00asse,bool00geasse,bool01gese,
bool02cr,bool03pse,christova,bool04geps,bool05pgese,
bool05geass,bool05geass2,bool05press,sharma} have suggested that there could be
not one but \textit{two} phase transitions near $\ra=2.4$, with the
opening of an \textit{intermediate phase} between the phases already
known.  The properties of this phase suggest that the intermediate
phase is rigid but stress-free. To explain the presence of two
transitions and the intermediate phase between them within the
framework of rigidity theory, it has been proposed \cite{thorpe00} that the glass
networks \textit{self-organize} in some way.  Broadly speaking, any
reduction in the amount of disorder in the network as it tries to
minimize its free energy can be referred to as self-organization;
chemical order, especially strong in oxide glasses like silica, would
be one example. Thorpe \textit{et al.}  \cite{thorpe00,czech01}
considered a particular kind of self-organization: glass networks
minimizing their elastic stress energy.  As a proof of principle, they
constructed the following model. Starting with a low-coordination
stress-free network, bonds are added one at a time with the restriction
that they cannot be redundant and thus add stress to the network,
using the pebble game for constructing and analyzing the network at
each step. This process is repeated until it is no longer possible to
add a bond without introducing stress to the network. After that, bond
insertion continues at random.  The maximum coordination at which no
stress is present, according to Eq.~(\ref{Maxexact}), cannot exceed
the rigidity threshold according to Maxwell counting ($\ra=4$ for 2D
CF networks and $\ra=2.4$ for covalent glass networks). In fact, in
the model of Thorpe \textit{et al.}, the Maxwell counting threshold value
is reached without stress for 2D CF networks, but not for the 3D glass
network \cite{czech01}. In this model, once the stress appears, it
immediately percolates, corresponding to the upper boundary of the
intermediate phase. In general, this need not be the case, since a
network can have finite stressed regions without stress percolating
(this is the case in the model of Micoulaut and Phillips
\cite{micoulaut02,micoulaut03}).

\begin{figure}
\begin{center}
\epsfig{file=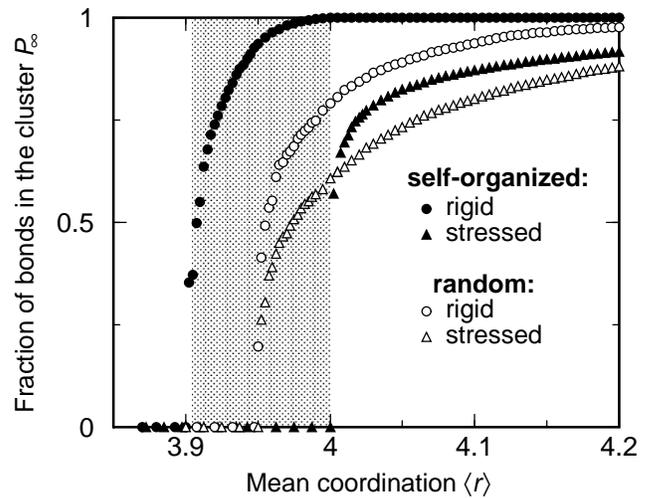,width=3.3in}
\caption{The dependence of the fraction of bonds in the percolating rigid
cluster and the percolating stressed region for the random case and the
self-organized case without equilibration. In the random case, the rigidity
and stress percolation thresholds coincide. In the self-organized case, these
thresholds do not coincide and there is an intermediate phase (shaded) between
them. The stressed region is defined here as the contiguous percolating set
of stressed bonds within the percolating rigid cluster.
The simulations are averaged over 2 realizations on the bond-diluted triangular
lattice of $400\times 400$ sites.
The figure is taken from Ref. \protect\cite{czech01}.\label{old}}
\end{center}
\end{figure}

To observe the rigidity transition in a
two-dimensional diluted regular lattice, one needs a lattice with
coordination bigger than 4, and so the triangular lattice is a natural
choice. Since the pebble game algorithm can only be applied to
generic networks, one has to assume that the triangular lattice is
distorted (for example, by having some disorder in bond
lengths). Previous numerical studies for the randomly diluted
triangular lattice without self-organization indicate a single
rigidity and stress transition at $\ra=3.961$ --- very close to the
Maxwell counting prediction (Fig.~\ref{old}) \cite{jacobs95}.  Note
that the rigidity and stress transitions coincide in this case.

In the self-organization model of Thorpe \textit{et al.}, a numerical
simulation reveals instead two phase transitions (Fig.~\ref{old}). As
the coordination is increased from the floppy phase, a percolating
rigid cluster appears at $\ra=3.905$. At this point, by construction,
the network is still stress-free. This is the lower boundary of the
rigid but stress-free intermediate phase --- the rigidity percolation
transition.  As the mean coordination continues to grow, keeping the
network stress-free becomes impossible. At this point, which, as
mentioned above, is at $\ra=4$, the stress appears and immediately
percolates. This is the stress percolation transition, which is the
upper boundary of the intermediate phase.

\begin{figure}
\begin{center}
\epsfig{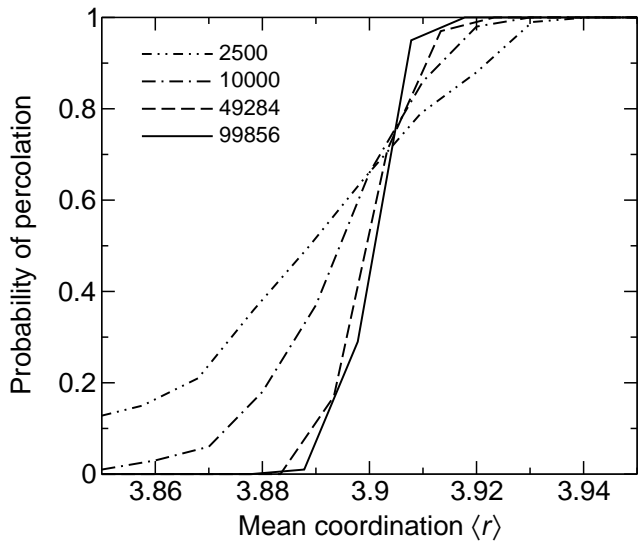}
\caption{The fraction of networks in which the percolating rigid cluster is
present as a function of $\ra$ for the model of self-organization without
equilibration. Each curve is obtained from 100 separate runs on a bond-diluted
triangular lattice; the lattice sizes are indicated in the legend.
\label{propercold}}
\end{center}
\end{figure}

The probability of a percolating cluster in the model of Thorpe \textit{
et al.} is zero below the rigidity percolation threshold and one above
in the thermodynamic limit, as normally is the case for percolation
transitions.  This is illustrated in Fig.~\ref{propercold}, where the
probability of having a percolating cluster is shown for different
network sizes. As expected, the dependence gets closer to a step
function as the size increases.

\section{A self-organization model with equilibration}

The self-organization model of Thorpe \textit{et al.} is peculiar in that
bonds are only added to the network and never removed. While one can
imagine a very rapid quench process in which indeed bond formation
dominates, this process does not lead to formation of good glasses.
Therefore a way of building equilibrated stress-free networks is
needed.  As the elastic energy of a stress-free network is zero,
any such networks should occur with equal probability.

A similar issue arose before in the case of conventional (or
connectivity) percolation. In the rigidity case, self-organization
proceeds by avoiding stress or redundancy, i.e., bonds that connect
already mutually rigid sites.  The connectivity analog consists in
avoiding connections between sites that are already connected, i.e.,
creating loops. Straley \cite{straley79} proposed a model directly
analogous to the one by Thorpe \textit{et al.}, i.e., with bonds inserted
one at a time and those forming loops rejected; connectivity
percolation occurs at some point, and then there is an ``intermediate
phase'' (although it was not referred to as such in the connectivity
percolation context) that is connected but without any loops, until a
point is reached at which avoiding loops is no longer possible (at
this point the network is a \textit{spanning tree}). It was realized
later on that this model does not produce a uniform ensemble, in which
every loopless network would occur with equal probability. Several
authors, using a variety of methods \cite{wu,family,straley90} ,
claimed then that in the equilibrated uniform ensemble, connectivity
percolation does not occur until the spanning tree limit is reached,
i.e., there is no intermediate phase. This shows that the results may
change significantly depending on the ensemble of self-organized
networks that is considered.

Braswell \textit{et al.} \cite{family}, in particular, used the following
algorithm to generate equiprobable loopless networks: take an
arbitrary loopless network, choose a bond at random, delete it and
then reinsert at an arbitrary place where it would not form a loop,
with this place chosen equiprobably among all such places. They showed
that after equilibration has taken place, this method would indeed
generate the uniform ensemble of networks, by proving the detailed
balance condition, i.e., that given some loopless network 1, the
probability that in a single step of the algorithm some other network
2 would be produced is the same as the probability of going in the
opposite direction, i.e., from network 2 to network 1. Their arguments
fully apply to the rigidity case as well.

In view of the above, we consider a variety of the self-organization
model by Thorpe \textit{et al.}, adding equilibration that produces
equiprobable stress-free networks. Like in the previous model, we start
with the ``empty'' network without bonds and start inserting bonds one
by one without creating stress. After every bond insertion, we
equilibrate by doing bond swaps following the procedure described
above, i.e., choose a bond at random, delete it and then insert a bond
elsewhere choosing at random among the places where that new bond
would not create stress. It is worth noting that in general it is 
rather difficult to handle removal of constraints within the pebble game; but
it is easy to remove an unstressed constraint, as this simply involves
releasing the associated pebble with no additional pebble
rearrangement. Since in our case all constraints are unstressed by
construction, no problem arises.  In this paper we focus on the
intermediate phase and do not investigate the stressed phase, so we
stop at the point at which further stressless insertion becomes
impossible.

A very similar model has been proposed recently by Barr\'e \textit{et
al.}  \cite{barre} using a Bethe lattice and, as an added
sophistication, an energy-cost function linear in the number of
redundant constraints. While this energy is somewhat unrealistic, it
provides a thermodynamic justification for the existence of the
intermediate phase. However, Bethe lattices are particular
constructions, leading to a first-order rigidity phase transition
while 2D central-force and 3D bond-bending networks show a
second-order rigidity transition. By comparison, our model in essence
assigns an infinite cost to redundant bonds and corresponds therefore
to the $T\to 0$ limit of Barr\'e's model but on a regular lattice with
a second-order rigidity transition.

\section{Rigidity percolation in the intermediate phase}

\begin{figure}
\begin{center}
\epsfig{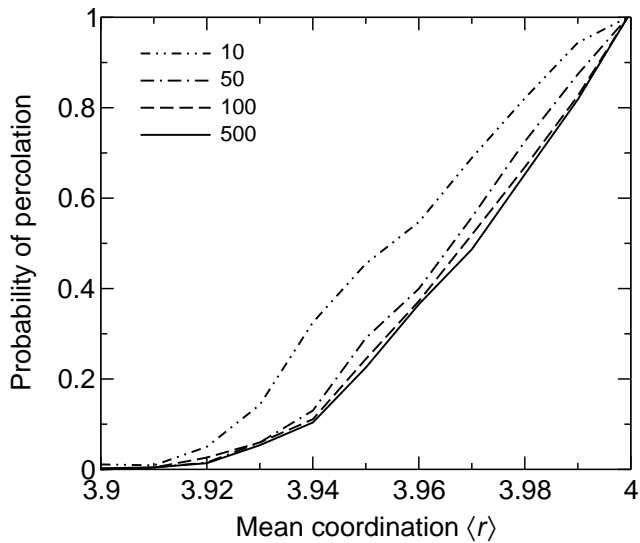}
\caption{Same as in Fig.~\ref{propercold}, but now for the model with
equilibration, for triangular networks of 10000 sites and for several
different equilibration times indicated in the legend as the number of
equilibration steps per each inserted bond above $\ra=3.5$; below
$\ra=3.5$, there are 10 equilibration steps per bond in all cases. The
data are likewise from 100 separate runs, but in addition, from each
run all networks obtained during the equilibration procedure at the
given mean coordination are taken into account.\label{properctime}}
\end{center}
\end{figure}

\begin{figure}
\begin{center}
\epsfig{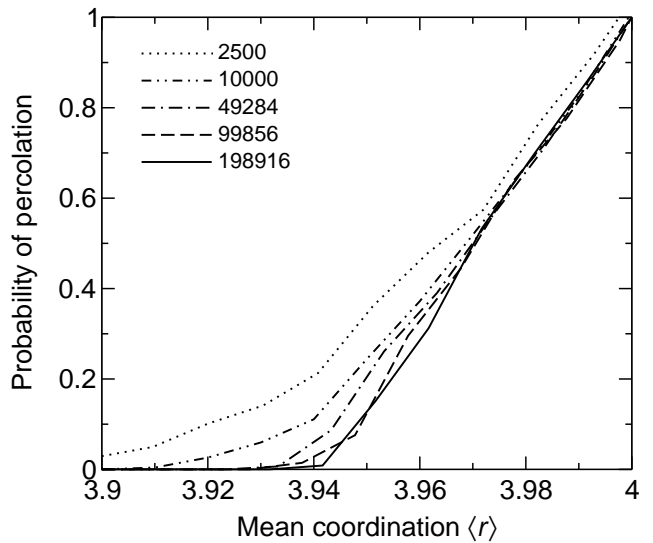}
\caption{Same as in Fig.~\ref{properctime}, but for different network sizes
indicated in the legend, always using 10 equilibration steps per inserted bond below
$\ra=3.5$ and 100 equilibration steps above.\label{propercsize}}
\end{center}
\end{figure}

Our simulations for the new model with equilibration are done for 2D
CF bond-diluted triangular lattices. Periodic boundary conditions were used
with the supercell consisting of the same number of unit cells in both
directions. We have chosen the duration of
equilibration equal to 100 steps above $\ra=3.5$ and 10 steps below
(where even in random networks there are very few redundant constraints, so the
self-organized networks are
almost completely random anyway and long equilibration is not
needed). This is sufficient for convergence as shown in Fig.~\ref{properctime}
for a $100\times100$ lattice; 
we also checked that 100 steps was sufficient by comparing with a very
long equilibration run at a single point $\ra=3.95$ for the largest lattice
used here (not shown).
The result we get is quite different from that obtained without
equilibration.  Fig.~\ref{propercsize}, just as Fig.~\ref{propercold},
shows the probability of rigidity percolation as a function of $\ra$ for
several different sizes, but now for the model with equilibration.
The self-organization still opens an intermediate
phase around the critical point found at $\ra\approx 3.961$ in the standard
rigidity phase diagram. But rather than approaching a step function as the size
increases, the result for the probability of percolation is now a gradual
increase from 0 at $\ra\approx 3.94$ to 1 at $\ra=4$.  This is similar to
the result on a Bethe
lattice in the model of Barr\'e \textit{et al.}, even though the
rigidity percolation transition is of a different order. The
dependence of the percolation probability on $\ra$ is close to linear
and this linearity may, in fact, be exact, although a very small
non-linear region near the lower boundary of the intermediate phase
cannot be ruled out.   One difference between our result and
that presented by Barr\'e \textit{et al.} is that since their
consideration is at a non-zero temperature, there is always a
possibility of having a small number of redundant constraints; since
adding very few (perhaps $\mathcal{O}(1)$) redundant constraints to an
unstressed network is often enough for stress percolation, it is not
surprising that they have found a finite probability of both rigidity
and stress percolation in the intermediate phase, whereas in our case,
the stress percolation probability is, of course, zero by
construction.

\section{Entropy cost of self-organization}

Although we do not include a potential energy explicitly, the self-organization
models discussed here are constructed to prevent the build-up of stress,
implicitly minimizing the potential energy of the system.
At a finite temperature, however, we are interested in minimizing the free
energy. As glasses are formed at a non-zero temperature, the thermodynamical
state will be influenced by the balance between the entropic cost associated
with generating a self-organized network and the energetic cost of creating
internal stress: if the entropy of the random network is large compared with
that of the self-organized one, then it is likely that very little
self-organization will take place and the problem discussed here becomes
irrelevant.

The entropy of a covalent network can be viewed as consisting of two
parts. The first part, the \textit{topological entropy}, is proportional
to the logarithm of the number of different possible bond topologies
or bond configurations. The second part, which can be called, somewhat
simplistically, the \textit{flexibility entropy}, depends on the phase
space available to each such bond configuration.  This division of the
total entropy is similar, but not identical, to the traditional
division into the configurational and vibrational entropy in the
inherent structure formalism \cite{stillinger}. In particular,
flexible networks exhibit a wide range of motions and would generally
correspond to more than one inherent structure, when a potential
energy function is defined. For this reason, the flexibility entropy
includes both harmonic and anharmonic contributions associated
with a given topology of the covalent network; its exact evaluation is
difficult and goes beyond the scope of this paper. However, since the
flexibility entropy is expected to be roughly proportional to the
number of floppy modes \cite{jacobsPRE,naumis,thorpe}, and, according to
Eq.~(\ref{Maxexact}), the number of floppy modes in a self-organized
network (with $N_R=0$) is smaller than in a random network ($N_R>0$)
with the same number of constraints $N_c$, the flexibility entropy of
the self-organized network is likely to be smaller than that of the
random network.  This difference is probably not very large,
especially in real systems where long-range forces reduce
significantly the available configuration space even in the floppy phase.

The topological entropy is, in general, difficult to calculate as
well, although methods, such as that by Vink and Barkema
\cite{barkema}, exist. It is much simpler for a lattice-based model
like ours, as it requires only counting the number of possible bond
configurations on a lattice. In this case, the topological entropy
(which we can also call the bond-configurational entropy) is simply
\begin{equation}
S_{\rm bc}(\ra)=\ln N_{\rm bc}(\ra),\label{entrdef}
\end{equation}
where $N_{\rm bc}(\ra)$ is the number of stress-free configurations with mean
coordination $\ra$ and the Boltzmann constant $k_B$ is put equal to 1. To
calculate $N_{\rm bc}$, we use the following approach. Suppose the number of
stress-free networks having $N_B$ bonds, $N_{\rm bc}(N_B)$, is known. From a
stress-free network having $N_B$ bonds, it is possible to produce a stress-free
network having $N_B+1$ bonds by adding a bond in one of those places where this
added bond would not create stress. Suppose on average there are $n_{+}(N_B)$
such places or ways to create a stress-free network with $N_B+1$ bonds. On the
other hand, for any stress-free network with $N_B+1$ bonds, there are always
exactly $n_{-}(N_B+1)=N_B+1$ ways of creating a stress-free network with $N_B$
bonds by removing any one of the $N_B+1$ bonds. Moreover, if a network with
$N_B+1$ bonds can be created from a network with $N_B$ bonds by adding a bond,
then the latter network can always be obtained from the former by removing that
same bond and vice versa, so the process is reversible. Then the number of
stress-free networks with $N_B+1$ bonds is
\begin{eqnarray}
N_{\rm bc}(N_B+1)&=&N_{\rm bc}(N_B)\cdot \frac{n_{+}(N_B)}{n_{-}(N_B+1)}\nonumber\\
&=&N_{\rm bc}(N_B)\cdot \frac{n_{+}(N_B)}{N_B+1}.
\label{iter}
\end{eqnarray}
If $n_{+}(N_B)$ is known for all $N_B$, then, using $N_{\rm bc}(0)=1$ as the
initial condition, Eq.~(\ref{iter}) can be iterated to yield all
$N_{\rm bc}(N_B)$. In practice, $n_{+}(N_B)$ are obtained numerically, by
a sort of Monte Carlo procedure, where some of the
places in the network where a bond is missing (compared to the full
undiluted lattice) are tried and it is found in what fraction of such places addition
of a bond would not create stress. In our simulations, we use 100 such attempts
per network. The result is then averaged over
the networks with a given number of bonds obtained during the equilibration
procedure.

Note that the same procedure can be repeated for the random case (without any
self-organization). In this case, $n_{+}(N_B)$ should count all ways to create a
new network of $N_B+1$ bonds (no matter stress-free or not) out of a network of
$N_B$ bonds, and there are as many ways to do that as there are places where a
bond is missing (compared to the full lattice); thus for a random network,
$n_{+}^{r}(N_B)=N_{Bf}-N_B$, where $N_{Bf}$ is the number of bonds in the full
lattice. On the other hand, the analog of quantity $n_{-}(N_B+1)$, which we
denote $n_{-}^{r}(N_B+1)$, is still $N_B+1$. Therefore,
for the random network,
\begin{eqnarray}
N_{\rm bc}^{r}(N_B+1)&=&N_{\rm bc}^{r}(N_B)\cdot\frac{n_{+}^{r}(N_B)}{n_{-}^{r}(N_B+1)}\nonumber\\
&=&N_{\rm bc}^{r}(N_B)\cdot\frac{N_{Bf}-N_B}{N_B+1}.
\end{eqnarray}
This can, of course, be used to obtain the analytical result for any $N_B$,
which is simply a binomial coefficient. If we are interested in the
\textit{difference} $\Delta S$ between the entropies of the random and
self-organized networks, this is simply
\begin{equation}
\Delta S(N_B)=\ln \frac{N_{\rm bc}^{r}(N_B)}{N_{\rm bc}(N_B)}.
\end{equation}
The change of this difference when a bond is added is
\begin{eqnarray}
\lefteqn{\Delta S(N_B+1)-\Delta S(N_B)=}\nonumber\\
&=&\ln \left(\frac{N_{\rm bc}^{r}(N_B+1)}
{N_{\rm bc}^{r}(N_B)}\cdot\frac{N_{\rm bc}(N_B)}{N_{\rm bc}(N_B+1)}\right)
\nonumber\\
&=&\ln\left(\frac{n_{+}^{r}(N_B)}{N_B+1}\cdot\frac{N_B+1}{n_{+}(N_B)}\right)
\nonumber\\
&=&-\ln\nu(N_B),
\label{s}
\end{eqnarray}
where $\nu(N_B)=n_{+}(N_B)/n_{+}^{r}(N_B)$ is the average \textit{fraction} of
bonds whose insertion would not create stress among all missing bonds in the
network. Note that $\nu(N_B)$ is what is calculated directly by the Monte
Carlo procedure described above.

The entropy $S$ is, of course, an extensive quantity, i.e., it is proportional
to the network size (for big sizes). We can introduce the entropy per bond of
the full lattice, $s=S/N_{Bf}$. In the thermodynamic limit,
\begin{eqnarray}
\frac{{\rm d}(\Delta s)}{{\rm d}\ra}&=&\frac{\Delta S(N_B+1)-\Delta S(N_B)}
{N_{Bf}\left[\ra\Big|_{N_B+1}-\ra\Big|_{N_B}\right]}\nonumber\\
&=&-\frac{\ln\nu}{N_{Bf}\left[\ra\Big|_{N_B+1}-\ra\Big|_{N_B}\right]},
\end{eqnarray}
where $\ra\Big|_{N_B}$ is the mean coordination of a network with $N_B$ bonds, and since
$\ra\Big|_{N_B}=2N_B/N$,
\begin{equation}
\frac{{\rm d}(\Delta s)}{{\rm d}\ra}=-\frac{N}{2N_{Bf}}\ln\nu.
\end{equation}
In the full triangular lattice, the number of bonds $N_{Bf}$ is three times the
number of sites $N$, so we get
\begin{equation}
\frac{{\rm d}(\Delta s)}{{\rm d}\ra}=-{1\over 6}\ln\nu.\label{deriv}
\end{equation}

\begin{figure}
\begin{center}
\epsfig{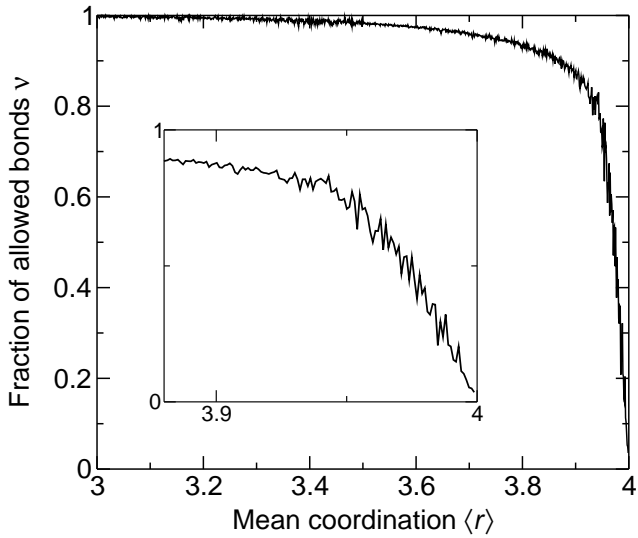}
\caption{The fraction $\nu$ of ``allowed bonds'' (places where a bond
can be inserted without creating stress) as a function of the mean
coordination $\ra$, for a triangular network of 50176 sites, with
equilibration time of 1000 steps per inserted bond above $\ra=3.5$ and
10 steps below. The result is obtained by a Monte Carlo procedure
described in the text; at each mean coordination, it is averaged over
the networks obtained during equilibration.\label{nu}}
\end{center}
\end{figure}

The quantity $\nu$ obtained numerically is plotted in Fig.~\ref{nu}. From this
plot, it seems to go to zero as $\delta=4-\ra\to 0^{+}$ and appears to change
linearly as a function of $\delta$ in this limit. In general, if in this limit
$\nu\sim\delta^m$, then, according to Eq.~(\ref{deriv}),
\begin{equation}
s(\delta)={m\over 6}\delta\ln\delta+\text{regular\ part}.\label{asymp}
\end{equation}

\begin{figure}
\begin{center}
\epsfig{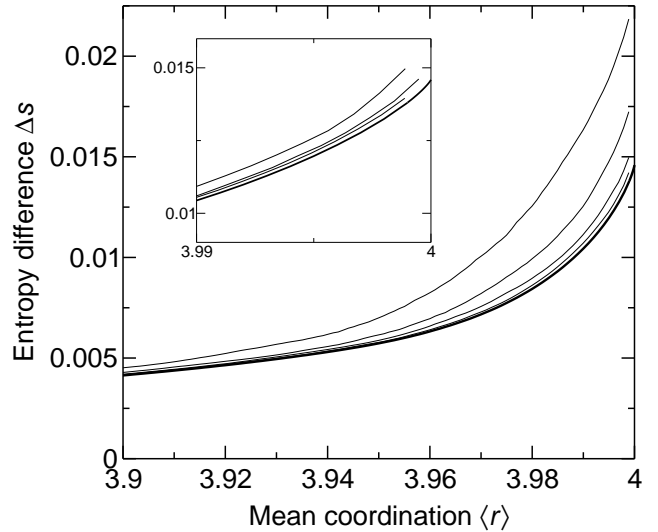}
\caption{The difference in the bond-configurational entropy between random and
  self-organized triangular networks, $\Delta s$, as a function of the mean
  coordination $\ra$, shown for several network sizes and equilibration times,
  as well as the asymptotic value (i.e., the extrapolation to the infinite size
  and relaxation time, as described in the text). Quantity $\nu$
  used to calculate $\Delta s$ is obtained by the Monte Carlo procedure
  described in the text and then averaged over networks obtained during equilibration
  at the given $\ra$. In the main plot, all thin
  lines are results for 50176 sites; the equilibration times are, top to bottom:
  10 steps, 30 steps, 100 steps and 300 steps per added bond. The thick line is
  the asymptotic. In the inset, all thin lines are results for 1000
  equilibration steps per added bond. The sizes are, top to bottom: 5476 sites,
  15376 sites and 50176 sites. The thick line is again the asymptotic. The
  asymptotics are themselves extrapolated to $\ra=4$, as described in the
  text. All equilibration times listed are for $\ra>3.5$; for $\ra<3.5$, 10
  equilibration steps per added bond are always used.
\label{entr}}
\end{center}
\end{figure}

Figure~\ref{entr} shows the entropy difference $\Delta s$ calculated by
iterating Eq.~(\ref{s}), with $\nu$ obtained numerically after every bond
addition. Since the simulations are done for networks of a finite size and
with finite equilibration time, an extrapolation to the infinite size was done
by fitting the simulation results to the function
\begin{equation}
\Delta s(\ra;N,\tau)={A(\ra)\over N}+{B(\ra)\over \tau}+C(\ra),
\end{equation}
where $\tau$ is the equilibration time (in equilibration steps per added bond)
used above $\ra=3.5$ (below 3.5, we always use just 10 steps, for reasons
explained above). As in different
runs for different sizes the data were taken at slightly different points,
linear interpolation was sometimes done to obtain the values at the same
$\ra$ in all cases.
In total, data for 182 $(N,\tau)$ combinations were used, with $N$ between
5476 and 50176 sites and $\tau$ between 20 and 1000 steps (up to 10000 steps
for 10000 sites). Data for higher $N$ and $\tau$ were assigned higher
weights in the fit, since increasing $N$ and $\tau$ decreases the amount of
noise in the data (the latter because the results for $\nu$ are averaged over
all equilibration steps, and thus the more equilibration steps there are the
smaller the error in $\nu$).
Function $C(\ra)$ representing the asymptotic value of the entropy difference
is also shown in Fig.~\ref{entr}.

For a finite-size network, it is only possible to reach a point 3 bonds short
of $\ra=4$ without creating stress. For this reason, our simulations were
stopped somewhere around $\ra=3.999$. Taking into account Eq.~(\ref{asymp}),
we have fitted the asymptotic entropy difference $C(\delta)$ between
$\ra=3.97$ and 3.999 using the following function:
\begin{equation}
C(\delta)=a_0+a_1 \delta\ln\delta+a_2 \delta+a_3 \delta^2.
\end{equation}
The fit is essentially perfect, and the obtained value of $a_1=0.1636$ is
consistent with the value of $1/6$ expected for $m=1$, according to
Eq.~(\ref{asymp}). The fit is used to complete the curve in Fig.~\ref{entr} up
to $\ra=4$. The value of the entropy difference at $\ra=4$ is $a_0=0.0146$.
This is the biggest value of the entropy difference, but it is still small, only
about 2\% of the bond-configurational entropy of the random network (which at
$\ra=4$, when 2/3 of the bonds are present, is
$-[(2/3)\ln (2/3)+(1/3)\ln (1/3)]=0.6365\ldots$).

In the above calculation of the entropy we explicitly use the fact that all
stress-free networks with a given number of bonds are equiprobable in our new
model. A similar consideration for the old self-organization model without
equilibration would be much more difficult.

We note, finally, that according to Eq.~(\ref{asymp}), there is a
non-analyticity in the behavior of the entropy when $\ra\to 4^{-}$, i.e., at the
stress transition. We should also expect some very weak non-analyticity (perhaps
a break or a cusp in a higher derivative) at the lower boundary of the
intermediate phase (the rigidity transition) at $\ra\approx 3.94$, but it is too weak
to be seen in our simulation results.

\section{Conclusion}

We have considered a model of self-organization in elastic networks,
adding an equilibration feature to the model previously considered by
Thorpe \textit{et al.}  \cite{thorpe00,czech01}. In our model, we find an
intermediate phase in the rigidity phase diagram, in which the
fraction of networks in which rigidity percolates is between 0
and 1 in the range of mean coordination between 3.94 and 4.0 for
the bond-diluted triangular lattice, a result qualitatively similar to
that obtained by Barr\'e \textit{et al.} \cite{barre} in a closely
related model on a Bethe lattice.

Calculating the bond-configurational entropy of these self-organized networks,
we find that it is only about 2\% smaller than that of randomly-connected
networks.  Provided that the flexibility entropy, which should reduce the
stability of the intermediate phase, is not so sensitive to the
self-organization, the intermediate phase is likely to be present in most
systems with the right range of mean coordination.

Our results support the current explanation of the intermediate phase in
chalcogenide glasses. Self-organization might also be
important in the dynamics of proteins as they have a coordination near the
critical value.  In a real material, it is likely that the intermediate phase is
not perfectly stress-free. A most likely structure will therefore be mostly
unstressed with overconstrained local regions, in a mixture of the model
presented here and that introduced recently by Micoulaut \textit{et al.}
\cite{micoulaut02,micoulaut03}.

The fact that there is now a possibility of having non-rigid networks
in the intermediate phase does not invalidate the concept of this
phase as lacking both excessive flexibility and stress. Indeed, even
though the network may technically be floppy, say, because, of a
single floppy mode or a very small number of such modes spanning the
whole network, for any practical purposes, there would be no
difference between such a network and a fully rigid and unstressed
one, especially when the existence of the neglected weaker
interactions is taken into account.

Knowing that self-organization can exist from a thermodynamical point of view,
there is still considerable work to do in order to fully understand the
intermediate phase. Among the obvious future directions of this work we can
mention: repeating the simulations described here for 3D bond-bending glass
networks; getting a better idea of the geometry of self-organized networks, in
particular, possible long-range correlations in
them; evaluating the flexibility entropy effects using a particular potential
energy function.

Finally, there is a suspicious discrepancy between our results
reported here and those obtained for a similar model in connectivity percolation
\cite{wu,family,straley90}. Even though it is in principle possible that
there is an intermediate phase in the rigidity case but not in the
connectivity case, this seems very unlikely. Perhaps the time has come
to re-evaluate these old results --- we certainly have the benefit of
the much increased computational power.

\begin{acknowledgments}
MC would like to thank M.F.~Thorpe for introducing him to this area of
research and many useful discussions.  MC acknowledges partial support
from the Minist\`ere de l'\'education, du sport et des loisirs
(Qu\'ebec). NM is supported in part by FQRNT (Qu\'ebec), NSERC and the
Canada Research Chair program. We thank the RQCHP for generous
allocation of computational resources.
\end{acknowledgments}

\end{document}